\documentclass[twoside,12pt]{article}
\usepackage{epsfig}

\def\Journal#1#2#3#4{{#1} {#2} (#4) #3 }

\def\NPA{{\em Nucl. Phys.} A}

\def\NPB{{\em Nucl. Phys.} B}

\def\PLB{{\em Phys. Lett.} B}

\def\PRL{\em Phys. Rev. Lett.}

\def\PRD{{\em Phys. Rev.} D}
\def\PRC{{\em Phys. Rev.} C}

\newcommand{\be}{\begin{equation}}
\newcommand{\ee}{\end{equation}}
\newcommand{\bea}{\begin{eqnarray}}
\newcommand{\eea}{\end{eqnarray}}
\newcommand{\nn}{\nonumber}
\begin{document}

\title{ \vspace{1cm} QCD equation of state: \\ 
Physical quark masses and asymptotic temperatures}
\author{M.\ Bluhm$^1$, B.\ K\"ampfer$^{1,2}$, R.\ Schulze$^1$, D.\
Seipt$^1$\\ 
\\
$^1$Forschungszentrum Dresden-Rossendorf, PF 510119, 01314 Dresden, Germany\\
$^2$Institut f\"ur Theoretische Physik, TU Dresden, 01062 Dresden, Germany}
\date{~\vspace{-1cm}}
\maketitle
\begin{abstract} 
Within a phenomenological quasiparticle model, the quark mass and 
temperature dependence of the QCD equation of state is discussed and 
compared with lattice QCD results. Different approximations for the 
quasiparticle dispersion relations are employed, scaling properties 
of the equation of state with quark mass and deconfinement temperature 
are investigated and a continuation to asymptotically large 
temperatures is presented. 
\end{abstract}
\section{Introduction \label{sec.1}}

Within the past years, physicists aimed at revealing the very 
nature of strongly interacting matter experimentally by 
ultra-relativistic heavy-ion collisions performed at 
CERN-SPS and BNL-RHIC (see e.~g.~\cite{CERNReport,RHICReport}). At the 
large local energy densities reached during the collision process, 
a new state of deconfined matter, dubbed quark-gluon plasma (QGP), 
is thought to be created. Originally, motivated by asymptotic 
freedom in QCD, the QGP was considered as a weakly interacting gas 
of quarks ($q$) and gluons ($g$). However, the success of 
hydrodynamical concepts (supplemented by assuming fast 
thermalization~\cite{Heinz03,Shuryak04,Shuryak05,Gyulassy05} and 
low viscosity~\cite{Teaney01,Kolb01,Csernai06}) in describing 
experimental results points to the necessity of viewing the QGP 
rather as a strongly coupled system. 

For hydrodynamic considerations, the QCD equation of state (EoS), 
which is related to the grand potential $\Omega$, is of paramount 
interest. In addition, the EoS is of significant importance for 
the description of the dynamics of the early universe or of compact 
stellar objects. In general, $\Omega$ depends on parameters being 
specific for QCD like quark masses $m_q$, flavor number $N_f$ or 
color number $N_c$ as well as on external conditions described by 
temperature $T$ and various chemical potentials $\{\mu_q\}$. From the 
theoretical side, much progress has been achieved in first-principle 
(lattice) QCD evaluations of the grand potential. Previously, the latter 
were technically restricted to too large $m_q$ values of 
$\mathcal{O}(T)$ translating into a pion mass of about 770 
MeV~\cite{Karsch03} at the deconfinement critical temperature $T_c$. 
Only recently, smaller quark mass parameters were 
considered~\cite{Cheng08} pushing the pion mass to about 
215 MeV~\cite{Gupta08,Soeldner08}. Nonetheless, despite the necessary 
continuum extrapolation of lattice QCD results, also an extrapolation 
to physical quark masses remains an inevitable step towards obtaining 
reliable results. 

A variety of phenomenological approaches describes fairly well 
bulk information found in lattice QCD calculations by adjusting 
suitable parameters. Among these, effective quasiparticle 
models~\cite{Peshier,Bluhm,Biro07,Ivanov05,Khvorostukin06,Cassing}, 
PNJL models~\cite{Ratti06,Roessner07,Mukherjee07}, (Polyakov-) quark-meson 
models~\cite{Schaefer05,Schaefer07} and a colored bound state 
model~\cite{ShuryakZahed04,Gelman06} have to be mentioned. Here, we 
employ a quasiparticle model (QPM) for the description of QCD 
thermodynamics. The quark mass dependence is directly implemented 
in the quasiparticle dispersion relations, which allows for a 
comparison with lattice QCD results employing different quark 
mass values and enables an extrapolation to the physical limit. 
The quark mass dependence of QCD excitations in one-loop 
approximation was investigated in~\cite{Seipt08}. Armed 
by dispersion relations motivated from these considerations, the quark 
mass dependence of $\Omega$ is also discussed here. 

Soon, even larger local energy densities may be reached at CERN-LHC, 
which might provide deeper insights into the early formation dynamics 
of our universe. For pure SU(3) gauge theory, lattice QCD thermodynamics 
was recently studied up to temperatures 
$\sim 10^7\,T_c$~\cite{Fodor07}. At such asymptotically large temperatures, 
analytical attempts based on perturbative means~\cite{Blaizot01,Kajantie06} 
account fairly well for available lattice QCD results. Here, we discuss 
the EoS at asymptotically large temperatures within our QPM, both, 
for pure SU(3) and for $N_f=2+1$. (Nevertheless, for early universe 
studies also the influence of the heavy quark sector becomes 
important~\cite{Laine06}.) 

\section{Quasiparticle dispersion relations \label{sec.2}}

In the following, we concentrate on matter anti-matter symmetric systems, 
represented by a net baryon density $n_B=0$. In this case, the QPM rests 
on the entropy density $s=\sum_{i=g,q}s_i$ 
with 
\be
s_i(T) = 2\epsilon_i\frac{d_i}{\pi^2}\int_0^\infty dk k^2 
\left(\ln \left[1+\epsilon_ie^{-\omega_i/T}\right] + 
\epsilon_i\frac{\omega_i/T}{e^{\omega_i/T}+\epsilon_i} \right) \,,
\label{eq.1}
\ee
where $d_q=N_fN_c$, $d_g=N_c^2-1$, $\epsilon_q=1$, $\epsilon_g=-1$ 
(for details cf.~\cite{Bluhm,Bluhm08}). This 
ansatz assumes that the thermodynamically relevant excitations at momenta 
$k\sim T$ are transverse gluons and regular quark modes. The 
pressure $p=-\Omega /V$ ($V$ is the volume) 
and other related thermodynamic quantities follow from 
integrating $s=dp/dT$. The quasiparticle dispersion relations entering 
Eq.~(\ref{eq.1}) can be represented by $\omega_{g,q}^2=k^2+\Pi_{g,q}(T)$, 
where $\Pi_{g,q}(T)$ denote temperature dependent self-energies. 
The self-energies can phenomenologically be approximated by 
\bea
\label{eq.2}
\Pi_g & = & \frac16 G^2T^2\left(N_c+\frac{N_f}{2}\right) \,, \\
\label{eq.3}
\Pi_q & = & m_q^2 + 2 m_f^2 + \alpha m_q m_f
\eea
with $m_f^2=(N_c^2-1)G^2T^2/(16N_c)$. These expressions 
(cf.~\cite{Bluhm,Bluhm08} for $\alpha=2$) are based on one-loop approximations 
to lowest order in $m_q$ in the asymptotic momentum region, where 
Eq.~(\ref{eq.3}) is supplemented by a bi-linear term $\alpha m_q m_f$ 
for including nonzero quark masses according to~\cite{Pisarski89}. $G^2$, 
replacing the QCD running coupling $g^2$, represents an effective coupling 
strength parametrized by 
\be
\label{eq.4}
G^2(T) = \left\{
    \begin{array}{l}
      \!\! G^2_{\rm (2-loop)} (\zeta(T)), \quad T{\,\ge\,}T_c,
      \\[3mm]
      \!\! G^2_{\rm (2-loop)}(\zeta(T_c)) + b \left(1{-}\frac{T}{T_c}\right) 
+ a \left(1{-}\frac{T}{T_c}\right)^2 , \ 
      T{\,<\,}T_c 
    \end{array}
  \right.
\ee
with $\zeta(T)=\lambda(T-T_s)/T_c$ which approaches the perturbative 
region at large $T$ in line with the two-loop expression of $g^2$. 
In $\zeta(T)$, $\lambda$ can be related to $\Lambda_{\rm QCD}$ while 
$T_s$ regulates $G^2$ near $T_c$. For $T < T_c$, $G^2$ changes 
drastically its behavior as dictated by lattice QCD results. 

Recently~\cite{Seipt08}, the quark mass dependence of thermal QCD 
excitations in the one-loop approximation was examined in some detail. 
The gauge invariant expression for the transverse gluon 
self-energy at asymptotic momenta was found as 
\bea
\label{eq.5}
\Pi_g & = & \frac16 G^2 T^2 \left(N_c+\frac12 \sum_{q=1}^{N_f} 
\mathcal{I}\left(\frac{m_q}{T}\right)\right) \,, \\
\label{eq.6}
\mathcal{I}(m_q/T) & = & \frac{12}{\pi^2} 
\left(\frac{m_q}{T}\right)^2 \int_0^\infty d \sigma 
\frac{\sigma^2}{\sqrt{1+\sigma^2}}
\frac{1}{(1+e^{\sqrt{1+\sigma^2}m_q/T})} \,.
\eea
The power expansion of $\mathcal{I}(m_q/T)$ for small $m_q/T$ involves a term 
resembling chiral logarithms, cf.~\cite{Seipt08}. 
Neglecting any quark mass dependence by setting $m_q/T\rightarrow 0$, one gets 
$\mathcal{I}(m_q/T)\rightarrow 1$ and thus Eq.~(\ref{eq.2}) is reproduced. 
This approximation is supported to some extent by the one-loop 
results~\cite{Seipt08}, 
as the energy of transverse gluon excitations increases only by a tiny 
amount when decreasing $m_q$. 

In the case of regular quark excitations, 
$\Pi_q=m_q^2 + m_f^2 \mathcal{F}_q(k,m_q)$ is found in~\cite{Seipt08}, 
where $\mathcal{F}_q$ (depending non-trivially on $k$ and $m_q$) encodes 
the effects induced by the thermal medium. Straightforward evaluation of this 
expression in the asymptotic momentum region yields 
$\Pi_q=m_q^2 + 2M_+^2$ with $M_+^2=\frac13 m_f^2(\mathcal{I}(m_q/T)+2)$ 
and $\mathcal{I}(m_q/T)$ given in Eq.~(\ref{eq.6}). Thus, in the limit 
$m_q\rightarrow 0$, $M_+^2$ reduces to $m_f^2$. Obviously and in contrast 
to Eq.~(\ref{eq.3}), a bi-linear term relating $m_q$ and $m_f$ is not 
present in this result. 
However, for momenta $k\sim T$ and small $m_q$, a bi-linear term can 
be motivated from the one-loop approximations. In general, one can 
represent $\mathcal{F}_q(k,m_q)=\alpha(k,m_q)\mathcal{F}_q(k=0,m_q)$, 
where, for small $m_q$ (and fixed $k$), $\alpha$ is rather independent 
of $m_q$ and a number between 1 and 2. Approximating 
$\mathcal{F}_q(k=0,m_q)$ for small but nonzero $m_q$ and small $g$ in 
line with a generalization of~\cite{Pisarski89}, 
\be
\label{eq.7}
\Pi_q = m_q^2 + \alpha\left(-\frac12m_q^2 + m_qM_+ + M_+^2\right)
\ee
is obtained for an approximation of the asymptotic quark self-energy. 
With this ansatz, the thermal quark mass effectively decreases with 
decreasing quark mass parameter, as in the case of Eq.~(\ref{eq.3}). 
In fact, both approximations Eq.~(\ref{eq.3}) and Eq.~(\ref{eq.7}) 
yield the same expression for the quark self-energy in the 
limit $m\ll 1$ and for $\alpha=2$ reading $\Pi_q=2m_qm_f + 2m_f^2$. 

\section{Quark mass extrapolation \label{sec.3}}

Benchmark of the considerations is the scaled entropy density 
$s/T^3$ as a function of $T/T_c$ for $N_f=2+1$. In~\cite{Karsch03}, 
a continuum estimate for $s/T^3$ is given by reporting lattice 
QCD results for the energy density $e$ and the interaction measure 
$\Delta\equiv(e-3p)$, where $s/T^3=(4e-\Delta)/(3T^4)$, for fairly large and 
temperature dependent quark mass parameters $m_{u,d}/T=0.4$ and 
$m_s/T=1$. Recently~\cite{Cheng08,Gupta08}, $s/T^3$ was calculated 
for much smaller quark masses. 
Corresponding to~\cite{DeTar07}, the quark mass parameters 
used in~\cite{Cheng08,Gupta08} can be 
approximated by $m_u/T=m_d/T=A_0/T^2+A_1/T+A_2$ with 
$A_0=0.298\cdot 10^{-3}$ GeV$^2$, $A_1=1.664\cdot 10^{-3}$ GeV 
and $A_2=0.002$ for $T$ given in GeV and $m_s=10\,m_u$. 
As $s/T^3$ in~\cite{Gupta08} is given as a function of $T$ in MeV, 
we scale by $T_c=190$ MeV in line with~\cite{Gupta08}, where 
$T_c=(190\pm 5)$ MeV is reported giving rise to an estimated error 
in $T/T_c$ according to $\Delta T_c=5$ MeV. Despite, $T_c$ is 
assumed to be approximately quark (and related pion) mass independent 
(cf.~\cite{Fraga08} and references therein). 

In the following, the QPM armed by the two different approximations 
for the dispersion relations discussed in Sec.~\ref{sec.2} is adjusted 
to the lattice QCD results from~\cite{Karsch03}. 
The application of Eqs.~(\ref{eq.2}) and~(\ref{eq.3}) 
in the quasiparticle dispersion relations is denoted by ''Fit 1'', 
while the use of Eqs.~(\ref{eq.5})-(\ref{eq.7}) is associated with 
''Fit 2''. Then, by 
extrapolating to the mass set-up employed in~\cite{Gupta08}, the proper 
implementation of the quark mass dependence in the QPM can directly 
be tested. For simplicity, any conceivable $m_q$ dependence in $G^2$ is 
naively neglected. 

\begin{figure}[t]
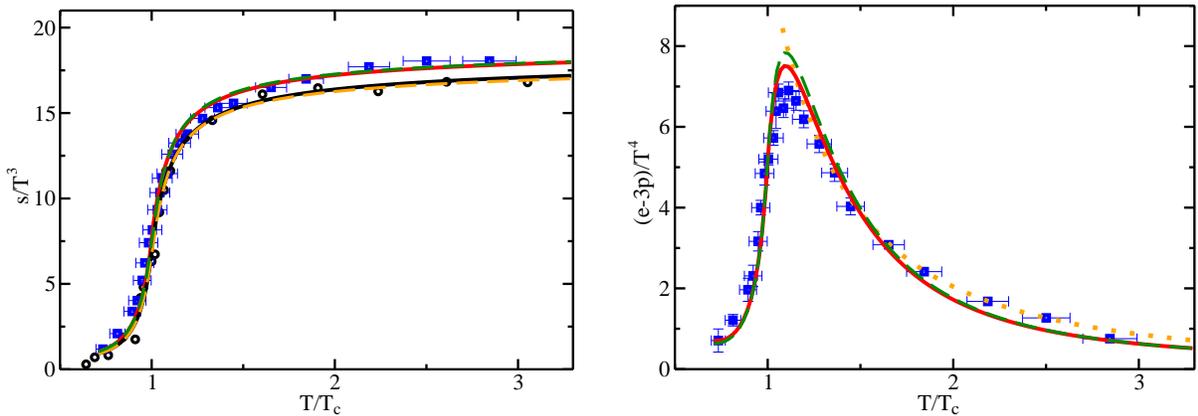

\begin{center}
\begin{minipage}[t]{16.5 cm}
\epsfig{file=entr19_18.eps,scale=0.32}
\hspace{4mm}
\epsfig{file=intmessure5.eps,scale=0.32}
\end{minipage}
\begin{minipage}[t]{16.5 cm}
\vspace{-.1cm}
\caption{Left: scaled entropy density $s/T^3$ as a function of 
$T/T_c$ for $N_f=2+1$. Circles~\cite{Karsch03} and squares~\cite{Gupta08} 
exhibit lattice QCD results for different quark mass set-ups. Dashed and 
solid curves depict QPM results for Fit 1 and Fit 2, respectively, 
employing different approximations for the quasiparticle dispersion 
relations as defined in Sec.~\ref{sec.2}. Lower curves represent adjustments 
to~\cite{Karsch03} while upper curves represent quark mass extrapolations 
in line with~\cite{Gupta08}. Right: scaled interaction measure 
$\Delta /T^4\equiv (e-3p)/T^4$ as a function of $T/T_c$. Squares denote 
lattice QCD results from~\cite{Gupta08,Soeldner08}, dashed and solid curves 
corresponding QPM results for Fit 1 and Fit 2, respectively. Dotted curve 
depicts the fuzzy bag model~\cite{Pisarski07} result for $T > 1.08\,T_c$. 
\label{fig1}}
\end{minipage}
\end{center}
\end{figure}

Starting with Fit 1, i.~e.~Eqs.~(\ref{eq.2}) and~(\ref{eq.3}) as 
approximation of the self-energies, 
we choose $\alpha=1$, here, as reasonable value 
(cf.~Sec.~\ref{sec.2}). The corresponding QPM parameters of $G^2$ read 
$T_s=0.7\,T_c$, $\lambda=5$, $a=-426$, $b=403.3$ describing $s/T^3$ 
from~\cite{Karsch03} impressively well as exhibited by the lower dashed 
curve in Fig.~\ref{fig1} (left panel). 
Extrapolating to smaller quark masses as used in~\cite{Gupta08}, the 
corresponding lattice QCD results are fairly well reproduced, cf.~upper 
dashed curve in Fig.~\ref{fig1} (left panel). Decreasing $m_q$, the 
thermal gluon mass remains unaffected, while the thermal quark mass is 
reduced according to Eq.~(\ref{eq.3}). Thus, the entropy density 
increases with decreasing quark mass. 

Note that an inclusion of the term $\alpha m_q m_f$ of significant 
strength seems to be mandatory. Neglecting this contribution by 
setting $\alpha =0$, any adjustment of QPM parameters to~\cite{Karsch03} 
fails in describing the quark mass extrapolation to~\cite{Gupta08}. 
This highlights the crucial role of the bi-linear term for the 
quark mass dependence of the phenomenological quasiparticle dispersion relations. 
For larger $\alpha$, $G^2$ has to take a smaller value at fixed $T$ 
for describing~\cite{Karsch03}. However, $\alpha$ cannot be chosen 
arbitrarily large, as a dominance of $\alpha m_q m_f$ might lead to 
an overestimation of the lattice QCD results~\cite{Gupta08} when 
extrapolating to the according quark mass values. We mention that 
using, instead, constant quark mass parameters $m_u=m_d=0$ MeV and 
$m_s=55 - 90$ MeV (which might be considered as physical limit) 
does not noticeably change the upper dashed curve on the scale 
exhibited in the left panel of Fig.~\ref{fig1}. 

In the case of Fit 2, i.~e.~Eqs.~(\ref{eq.5}) -~(\ref{eq.7}) as ansatz for the 
self-energies, we first note that the thermal gluon mass 
increases with decreasing $m_q$ according to Eq.~(\ref{eq.5}), 
while the behavior of the thermal quark mass is, in general, influenced by 
two counter-acting effects: $M_+$ increases with decreasing $m_q$ 
whereas the terms solely $\propto m_q$ decrease. Thus, depending on 
an appropriate choice for $\alpha$, $\Pi_q$ might decrease 
with decreasing $m_q$ mostly as a result of $\alpha m_q M_+$. 
Consequently, a larger value for $\alpha$ has to be expected 
compared to Fit 1 in order to saturate the quark mass behavior of 
$s/T^3$ from~\cite{Gupta08}. Here, for example, we choose $\alpha=2$. 
The corresponding QPM parameters, being similar to Fit 1, read 
$T_s=0.68\,T_c$, $\lambda=4.77$, $a=-426$ and $b=403.3$ when 
adjusting to $s/T^3$ from~\cite{Karsch03}, cf.~lower solid curve 
in Fig.~\ref{fig1} (left panel). The extrapolation to smaller quark 
masses in line with~\cite{Gupta08} is exhibited by the upper solid curve 
in the left panel of Fig.~\ref{fig1}. In both cases, Fit 1 and Fit 2 
(on the given scale solid and dashed curves lie almost on top of each 
other in Fig.~\ref{fig1} - left panel), 
agreement with the corresponding lattice QCD results is found. 

In addition, in~\cite{Gupta08,Soeldner08}, the scaled interaction measure 
$\Delta /T^4$ was calculated. The QPM results, according to Fit 1 and Fit 2 
for the small quark mass set-up, are depicted by dashed and solid curves in 
the right panel of Fig.~\ref{fig1}, respectively, finding an overall 
good agreement. The necessary pressure integration constant~\cite{Bluhm}, 
denoted by $B(T_c)$, reads $B(T_c)=0.65\,T_c^4$. Nonetheless, the QPM $m_q$ 
extrapolations overshoot somewhat the lattice QCD results around $T_c$ 
(Fit 1 more than Fit 2) and exhibit small deviations also at larger 
temperatures. 

Contrary, in a fuzzy bag model approach~\cite{Pisarski07}, lattice 
QCD results of $\Delta /T^4$ for $T > 1.2\,T_c$ can perfectly be 
described, cf.~dotted curve in Fig.~\ref{fig1} (right panel). The fuzzy 
bag picture originally accounts accurately 
for the plateau observed in the scaled 
interaction measure $\Delta /T^2$ as a function of $T$ (cf.~dotted 
curve in Fig.~\ref{fig2} - left panel), which is parametrized by 
$\Delta /T^2 = 2 B_f + 4 B_{MIT}/T^2$. Considering leading non-perturbative 
contributions to the pressure to be given by a temperature dependent 
(fuzzy) bag constant $B_f$ mimicking a gradual rather than an abrupt 
transition from the confined phase to a nearly perturbative phase, this 
picture represents a generalization of the MIT bag model. The fit 
parameters reproducing $\Delta /T^4$ for $T > 1.2\,T_c$ 
from~\cite{Gupta08,Soeldner08} in Fig.~\ref{fig1} read 
$B_f=0.135$ GeV$^2$, $B_{MIT}=0.0009$ GeV$^4$. Nonetheless, this 
approach neither describes the behavior of the interaction measure 
in the transition region nor incorporates explicitly 
quark mass effects. 

\section{Scaling properties of the equation of state \label{sec.4}}

Discussing the scaling properties of the QPM EoS with $m_q$ and the 
value of $T_c$, one first notes that the pressure reformulated in a 
dimensionless fashion reads 
\bea
\nn
    \Phi(\xi) \equiv \frac{p(T)}{T^4} & = & 
    \sum_{i=g,q} \epsilon_i \frac{d_i}{\pi^2} \int_0^\infty 
    dx x^2 \ln \left(1+\epsilon_i e^{-\sqrt{x^2+\tilde{\Pi}_i}}\right)
    - b_0 \xi^{-4} \\
\label{eq.8}
    & & \hspace{-3.8cm}
    + \sum_{i=g,q} \frac{d_i}{2\pi^2\xi^4} \int_{1}^{\xi} 
    \left(2\tilde{\Pi}_i'+\eta\frac{\partial\tilde{\Pi}_i'}{\partial\eta}\right) 
    \eta^3 
    \left(\int_0^\infty \frac{dx x^2}{\sqrt{x^2+\tilde{\Pi}_i'}}
    \frac{1}{e^{\sqrt{x^2+\tilde{\Pi}_i'}}+\epsilon_i}\right)
    d \eta 
\eea
with $\tilde{\Pi}_i=\Pi_i(T)/T^2$, $\tilde{\Pi}_i'=\Pi_i(T')/T'^2$, 
real number $b_0=B(T_c)/T_c^4$ and $\xi\equiv T/T_c$, $\eta\equiv T'/T_c$. 
Eq.~(\ref{eq.8}) depends explicitly on $m_q$ via $\Pi_i$. 
For $m_q\equiv\epsilon\,T$ as used in~\cite{Karsch03}, 
$\tilde{\Pi}_i$ ($\tilde{\Pi}_i'$) is a function of $\xi$ ($\eta$) only 
because the entering $G^2$ depends on $T_c$ only via $\xi$ ($\eta$) (note 
that $\lambda T_s/T_c$ is a parameter in the QPM). 
Thus, $\Phi(\xi)$ displayed as a function of $\xi$ is 
independent of the explicit value of $T_c$. Consequently, related 
thermodynamic quantities like $s/T^3$ and $\Delta /T^4$ show the same 
independence of $T_c$. Nonetheless, approximating $m_q$ as 
advocated above in line with~\cite{Cheng08,Gupta08}, $\tilde{\Pi}_q$ 
explicitly depends on $T$, and thus, $\Phi(\xi)$ does not exhibit 
the discussed $T_c$ independence. However, numerically even 
a variation in $T_c$ by 100 MeV turns out to imply negligible 
effects on $p/T^4$, $s/T^3$ or $\Delta /T^4$. 

\begin{figure}[t]
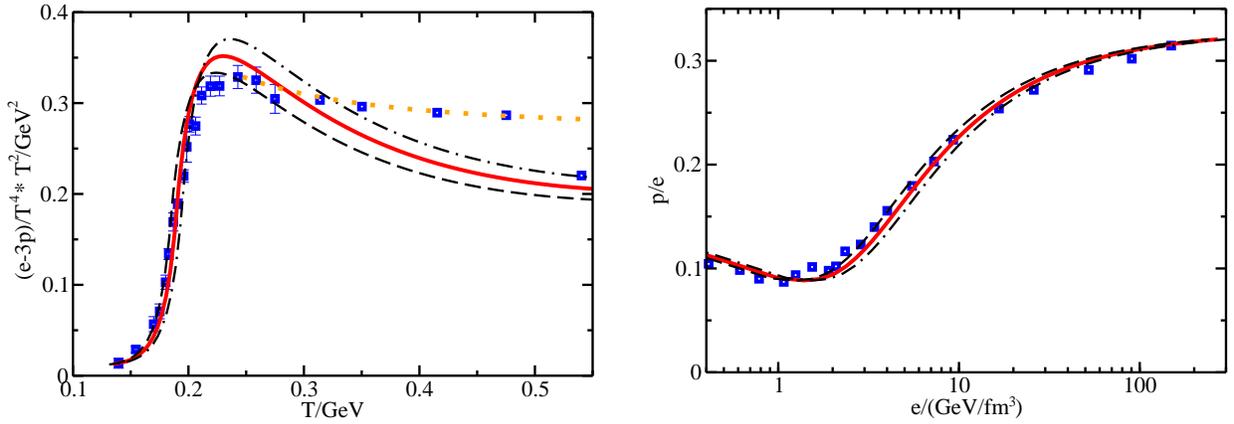

\begin{center}
\begin{minipage}[t]{16.5 cm}
\epsfig{file=intmessure8.eps,scale=0.32}
\hspace{4mm}
\epsfig{file=EoS2008_1.eps,scale=0.32}
\end{minipage}
\begin{minipage}[t]{16.5 cm}
\vspace{-.1cm}
\caption{Left: scaled interaction measure $(\Delta /T^2)/GeV^2$ as a function 
of T/GeV. Squares exhibit lattice QCD results from~\cite{Gupta08,Soeldner08}. 
Dashed, solid and dash-dotted curves depict QPM results according to 
Fit 2 using $T_c=185,\,190,\,195$ MeV, respectively. Dotted curve shows 
the directly fitted fuzzy bag model~\cite{Pisarski07} result. 
Right: EoS in the form $p/e$ as a function of $e/(GeV/fm^3)$ and its 
$T_c$ dependence. Lattice QCD results (squares) 
from~\cite{Schmidt08}, curves depict QPM results with the same line 
code as in the left panel. 
\label{fig2}}
\end{minipage}
\end{center}
\end{figure}

The interaction measure scaled by $T^2$ and depicted as 
a function of $T$ shows a $T_c$ dependence, cf.~Fig.~\ref{fig2} 
(left panel). A shift in $T_c$ by $5$ ($-5$) MeV in line 
with~\cite{Gupta08} results in a shift of the curve $\Delta /T^2$ to 
the right (left) in the transition region. In addition, 
the curves suffer a parallel shift up (down) by about 6\,-7\% for 
larger temperatures. Note that the deviations in $\Delta /T^4$ observed 
in Fig.~\ref{fig1} between QPM and~\cite{Gupta08,Soeldner08} 
for larger $T$ are transparently quantified by deviations from the 
plateau behavior in Fig.~\ref{fig2} (left panel). 

The parallel shift in $\Delta /T^2$ can qualitatively 
be understood from the scaling 
behavior of the EoS with $T_c$. Reformulating $p=T_c^4\xi(\Phi\xi^3)$ 
and imposing $\Phi(\xi)=\Phi(\xi')$, cf.~Eq.~(\ref{eq.8}), $p$ and 
thus $e=-p+\xi\partial p/\partial\xi$ change when changing $T_c$ to 
$T_c'$ (i.~e.~$\xi$ to $\xi'$) 
according to $p'=p (T_c'/T_c)^4$ and $e'=e (T_c'/T_c)^4$. 
For the EoS in the form $p(e)$, this implies that the linear section 
at larger $e$, which can be approximated by $p(e)=\alpha e+p_0$, becomes 
$p'(e')=\alpha e'+(T_c'/T_c)^4p_0$ with the same slope 
$\alpha$ but different off-set $p_0'=(T_c'/T_c)^4p_0$. 
For $T_c'>T_c$, the linear section of $p(e)$ is, thus, 
parallely shifted downward whereas for $T_c'<T_c$ it is 
shifted upward. A similar behavior is observed for the EoS in the 
form $p/e$ as exhibited in Fig.~\ref{fig2} (right panel). 

Other regions of $p(e)$ might need to be approximated 
differently, say for instance, by $p(e)=\tilde{\alpha}\sqrt{e}+\tilde{p_0}$ 
close to the transition region. 
Changing again $T_c$ to $T_c'$, 
the EoS changes into 
$p'(e')=\tilde{\alpha}'\sqrt{e'}+\tilde{p_0}'$ with 
$\tilde{\alpha}'=(T_c/T_c')^4\alpha$ and 
$\tilde{p_0}'=(T_c'/T_c)^4\tilde{p_0}$ implying a change 
in the off-set $\tilde{p_0}$ but also a flatter curve, 
$\tilde{\alpha}'<\tilde{\alpha}$, for $T_c'>T_c$ and a 
steeper curve, $\tilde{\alpha}'>\tilde{\alpha}$, 
for $T_c'<T_c$. This leads, now for $p/e$, to a change in the order 
of curves as evident from Fig.~\ref{fig2} (right panel) at lower 
energy densities. 

Likewise, one may discuss the $m_q$ dependence of the EoS. As empirically 
evident from Fig.~\ref{fig1}, the $m_q$ dependence of $s/T^3$ 
can be parametrized by 
$s'/T^3=\tilde{m}(\xi)s/T^3$ with $T$ dependent function 
$\tilde{m}$. Similarly, $p$ and $e$ (as discussed in~\cite{Bluhm08}) 
behave according to $p'=m(\xi)p$ and 
$e'=m(\xi)e + \xi p \partial m(\xi)/\partial\xi$ with $m_q$. Numerically, 
one finds $m(\xi)=1.107 \,...\, 1.081$ in the interval 
$\xi=1.5 \,...\, 3$, i.~e.~small changes in $m(\xi)$ with $\xi$, 
whereas for small $\xi$ a larger $m(0.8)=1.33$ and 
$m(0.9)=1.51$ is found. Thus, $m(\xi)$ varies more sizeably with 
$\xi$ for small $\xi$. 
Approximating $p(e)$ again by a linear function, 
$p'=\alpha e'/\left(1+\frac{\alpha\xi}{m}\frac{\partial 
m(\xi)}{\partial\xi}\right) + p_0 m/\left(1+
\frac{\alpha\xi}{m}\frac{\partial m(\xi)}{\partial\xi}\right)$ 
is found. Consequently, changes in $p(e)$ with $m_q$ are negligible 
for larger $\xi$, where $m(\xi)\approx 1$ is almost constant, 
whereas they are mostly visible in a region of $\xi$, 
where $m(\xi)$ varies most rapidly, i.~e.~for $\xi\le 1$. 

\section{Equation of state at asymptotic temperatures \label{sec.5}}

\begin{figure}[t]
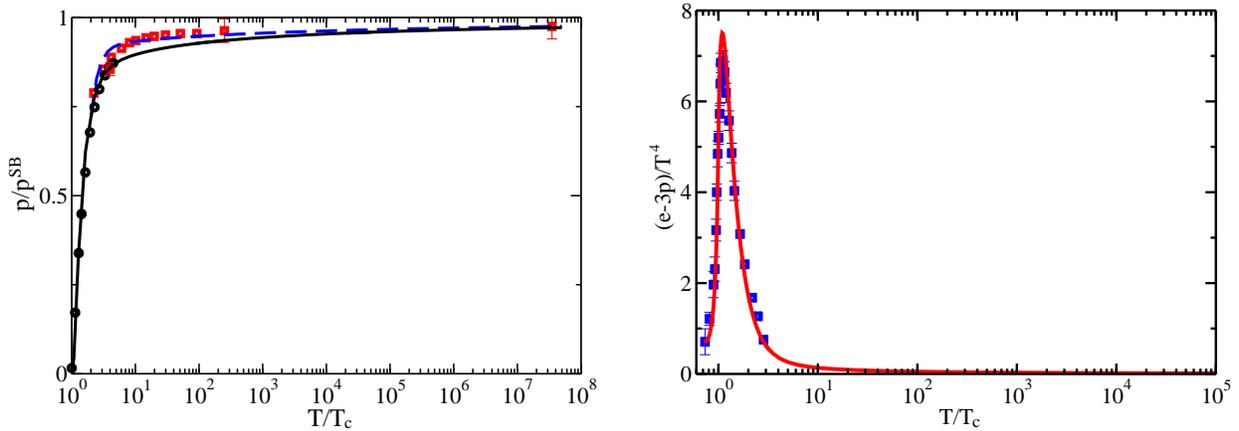

\begin{center}
\begin{minipage}[t]{16.5 cm}
\epsfig{file=pres14_4.eps,scale=0.32}
\hspace{4mm}
\epsfig{file=intmessure5_2.eps,scale=0.32}
\end{minipage}
\begin{minipage}[t]{16.5 cm}
\vspace{-.1cm}
\caption{Left: comparison of QPM (solid and dashed curves) with 
lattice QCD results (circles~\cite{Boyd96} and squares~\cite{Fodor07}) 
for $p/p^{SB}$ as a function of $T/T_c\ge 1$ of a pure gluon gas. 
The QPM is adjusted to~\cite{Boyd96} with $T_s=0.855\,T_c$, 
$\lambda=7$, $B(T_c)=-0.03\,T_c^4$ (solid curve) or 
to~\cite{Fodor07} with $T_s=0.97\,T_c$, 
$\lambda=150$, $B(T_c)=5\,T_c^4$ (dashed curve), where 
$T_c=271$ MeV~\cite{Karsch02}. Dashed curve is similar to 
perturbative QCD results reported in~\cite{Kajantie06,Laine06}. 
Right: scaled interaction measure (Fit 2 in Fig.~\ref{fig1}, right panel) 
as a function of $T/T_c$ continued to larger $T$. 
\label{fig3}}
\end{minipage}
\end{center}
\end{figure}

Recently~\cite{Fodor07}, the pressure for pure SU(3) gauge theory 
became available in a large temperature interval between $T_c$ 
and $10^7\,T_c$ confirming old results~\cite{Boyd96} between $T_c$ 
and $5\,T_c$. As $p$ changes most rapidly in the transition 
region, we fit the QPM parameters to~\cite{Boyd96} and then 
continue to larger $T$. The result for $p/p^{SB}$, where $p^{SB}$ 
denotes the Stefan-Boltzmann pressure, is exhibited by the solid 
curve in Fig.~\ref{fig3} (left panel). 
While the QPM describes appropriately the regions around $T_c$ and 
at asymptotically large temperatures, say for $T > 200\,T_c$, it 
underestimates~\cite{Fodor07} in the intermediate $T$ region between 
$3\,T_c$ and $200\,T_c$. Although one might find an exotic QPM 
parametrization for intermediate and large $T$ reproducing~\cite{Fodor07} 
and delivering similar results as perturbative QCD~\cite{Kajantie06,Laine06}, 
cf.~dashed curve in Fig.~\ref{fig3} (left panel), such a description 
fails for $T < 2\,T_c$. This discrepancy might be viewed as 
a hint for missing degrees of freedom in the intermediate $T$ region. 
Considering a hard-thermal-loop inspired extension of the 
QPM~\cite{Schulze}, including in addition longitudinal gluon (plasmon) 
and Landau damping contributions, the situation observed in the left 
panel of Fig.~\ref{fig3} is not measurably improved. 
This is due to the fact that 
the effect of a negative partial plasmon pressure decreasing for 
$T\rightarrow T_c^+$ is compensated by the interwoven increasing 
partial transverse gluon pressure (driven by the Landau damping) 
for $T\rightarrow T_c^+$, while 
the plasmon contribution vanishes for increasing $T$. Nonetheless, 
to resolve this issue, additional lattice QCD studies would be 
desirable. 

Turning to $N_f=2+1$ at asymptotic temperatures, the continuation 
of the scaled interaction measure (cf.~Fit 2 in Fig.~\ref{fig1} - 
right panel) to larger $T$ is exhibited in Fig.~\ref{fig3} (right 
panel). For large $T$, $\Delta /T^4$ approaches zero logarithmically 
in line with the temperature dependence of $G^2$. While already for 
$T\sim 10\,T_c$ the conformal limit $e=3p$ is approximately reached, 
$p/T^4$ still exhibits deviations from $p^{SB}/T^4$ of about 10\% 
at $T=50\,T_c$ and of 4.5\% at $T=10^5\,T_c$. As the 
dynamically generated thermal gluon 
and quark masses exhibit a behavior $\sim G T$ in the QPM, 
they are of the order of 12 TeV and 8 TeV, respectively, at 
$T=10^5\,T_c$, invalidating the naive picture of weakly coupled 
quarks and gluons with negligible masses. 

\section{Conclusion \label{sec.6}}

In summary, we study the quark mass and temperature dependence 
of the QCD equation of state. We utilize a quasiparticle model by 
employing two different expressions for the quasiparticle dispersion 
relations, which explicitly depend on $m_q$ and are based on 
one-loop QCD approximations~\cite{Seipt08}. In both cases remarkable 
agreement with first-principle lattice QCD results is achieved when 
extrapolating in the $m_q$ parameter space. Scaling properties 
of the QCD EoS with $m_q$ and $T_c$ are discussed and the EoS is 
continued to asymptotically large $T$, where very hot QCD matter may 
be viewed as composed of rather heavy quasiparticle excitations. 

\vspace{0.1cm}
\noindent
The authors thank E.~Laermann and Z.~Fodor for valuable discussions. 
The work is supported by BMBF 06DR136 and EU I3HP.


\begin{thebibliography}{99}
\itemsep -2pt 
\bibitem{CERNReport} C. H\"ohne (NA49 Collaboration), 
\Journal{\NPA} {774} {35} {2006}
\bibitem{RHICReport} {\it The First Three Years of Operation of RHIC}, 
\Journal{\NPA} {757} {1} {2005}
\bibitem{Heinz03} U. Heinz, \Journal{\NPA} {721} {30} {2003}
\bibitem{Shuryak04} E. V. Shuryak, {\it Prog. Part. Nucl. Phys.} 53 (2004) 273
\bibitem{Shuryak05} E. V. Shuryak, \Journal{\NPA} {750} {64} {2005}
\bibitem{Gyulassy05} M. Gyulassy, and L. D. McLerran, \Journal{\NPA} 
{750} {30} {2005}
\bibitem{Teaney01} D. Teaney, J. Lauret, and E. V. Shuryak, \Journal{\PRL} 
{86} {4783} {2001}
\bibitem{Kolb01} P. F. Kolb, P. Huovinen, U. Heinz, and H. Heiselberg, 
\Journal{\PLB} {500} {232} {2001}
\bibitem{Csernai06} L. P. Csernai, J. I. Kapusta, and L. D. McLerran, 
\Journal{\PRL} {97} {152303} {2006}
\bibitem{Karsch03} F. Karsch, K. Redlich, and A. Tawfik, {\it Eur. Phys. J.} C 
29 (2003) 549
\bibitem{Cheng08} M. Cheng et al., \Journal{\PRD} {77} {014511} {2008}
\bibitem{Gupta08} R. Gupta (HotQCD Collaboration), 
preprint arXiv:0810.1764 [hep-lat]
\bibitem{Soeldner08} W. S\"oldner (RBC-Bielefeld and HotQCD Collaborations), 
preprint arXiv:0810.2468 [hep-lat]
\bibitem{Peshier} A. Peshier, B. K\"ampfer, O. P. Pavlenko, and G. Soff, 
\Journal{\PLB} {337} {235} {1994}, \Journal{\PRD} {54} {2399} {1996}; 
A. Peshier, B. K\"ampfer, and G. Soff, \Journal{\PRC} {61} {045203} {2000}, 
\Journal{\PRD} {66} {094003} {2002}
\bibitem{Bluhm} M. Bluhm, B. K\"ampfer, and G. Soff, \Journal{\PLB} 
{620} {131} {2005}; M. Bluhm et al., \Journal{\PRC} {76} {034901} {2007}; 
M. Bluhm, B. K\"ampfer, R. Schulze, and D. Seipt, {\it Eur. Phys. J.} C 
49 (2007) 205
\bibitem{Biro07} T. S. Biro, P. Levai, P. Van, and J. Zimanyi, 
\Journal{\PRC} {75} {034910} {2007}
\bibitem{Ivanov05} Y. B. Ivanov et al., \Journal{\PRC} {72} {025804} {2005}
\bibitem{Khvorostukin06} A. S. Khvorostukin, V. V. Skokov, V. D. Toneev, and 
K. Redlich, {\it Eur. Phys. J.} C 48 (2006) 531
\bibitem{Cassing} W. Cassing, \Journal{\NPA} {791} {365} {2007}, 
\Journal{\NPA} {795} {70} {2007}
\bibitem{Ratti06} C. Ratti, M. A. Thaler, and W. Weise, \Journal{\PRD} 
{73} {014019} {2006}
\bibitem{Roessner07} S. R\"o\ss ner, C. Ratti, and W. Weise, \Journal{\PRD} 
{75} {034007} {2007}
\bibitem{Mukherjee07} S. Mukherjee, M. G. Mustafa, and R. Ray, \Journal{\PRD} 
{75} {094015} {2007}
\bibitem{Schaefer05} B.-J. Sch\"afer, and J. Wambach, \Journal{\NPA} 
{757} {479} {2005}
\bibitem{Schaefer07} B.-J. Sch\"afer, J. M. Pawlowski, and J. Wambach, 
\Journal{\PRD} {76} {074023} {2007}
\bibitem{ShuryakZahed04} E. V. Shuryak, and I. Zahed, \Journal{\PRD} 
{70} {054507} {2004}
\bibitem{Gelman06} B. A. Gelman, E. V. Shuryak, and I. Zahed, 
\Journal{\PRC} {74} {044908} {2006}, \Journal{\PRC} {74} {044909} {2006}
\bibitem{Seipt08} D. Seipt, M. Bluhm, and B. K\"ampfer, 
preprint arXiv:0810.3803 [hep-ph]
\bibitem{Fodor07} G. Endr\"odi, Z. Fodor, S. D. Katz, and K. K. Szabo, 
{\it Proc. Sci.} LAT2007 (2007) 228
\bibitem{Blaizot01} J.-P. Blaizot, A. Rebhan, and E. Iancu, 
\Journal{\PRD} {63} {065003} {2001}
\bibitem{Kajantie06} K. Kajantie, M. Laine, K. Rummukainen, and 
Y. Schr\"oder, \Journal{\PRD} {67} {105008} {2006}
\bibitem{Laine06} M. Laine, and Y. Schr\"oder, \Journal{\PRD} 
{73} {085009} {2006}
\bibitem{Bluhm08} M. Bluhm, and B. K\"ampfer, preprint arXiv:0807.4080 [hep-ph]
\bibitem{Pisarski89} R. D. Pisarski, \Journal{\NPA} {498} {423c} {1989}
\bibitem{DeTar07} C. DeTar, and R. Gupta (HotQCD Collaboration), 
{\it Proc. Sci.} LAT2007 (2007) 179
\bibitem{Fraga08} E. S. Fraga, L. F. Palhares, and C. Villavicencio, 
preprint arXiv:0810.1060 [hep-ph]
\bibitem{Pisarski07} R. D. Pisarski, {\it Prog. Theor. Phys. Suppl.} 
168 (2007) 276
\bibitem{Schmidt08} C. Schmidt (HotQCD Collaboration), 
preprint arXiv:0810.0374 [hep-lat]
\bibitem{Boyd96} G. Boyd et al., \Journal{\NPB} {469} {419} {1996}
\bibitem{Karsch02} F. Karsch, {\it Lect. Notes Phys.} 583 (2002) 209
\bibitem{Schulze} R. Schulze, M. Bluhm, and B. K\"ampfer, 
{\it Eur. Phys. J. ST} 155 (2008) 177, preprint arXiv:0803.1571 [hep-ph]; 
R. Schulze, and B. K\"ampfer, preprint arXiv:0811.0274 [hep-ph], these 
proceedings
\end{thebibliography}
\end{document}